\documentclass[journal]{IEEEtran}
\usepackage{graphicx}
\usepackage{multirow}

\usepackage{amsfonts}
\usepackage[numbers,sort&compress,square]{natbib}
\usepackage{siunitx}
\usepackage[pdf]{pstricks}
\ifCLASSINFOpdf
\else
\fi
\usepackage{lineno}
\usepackage{nomencl}
\makenomenclature
\usepackage{algorithm}
\usepackage{algorithmic}
\usepackage{amstext}
\usepackage{amsmath}
\usepackage{amsfonts}
\usepackage{color}
\usepackage{array}
\usepackage{nomencl}
\usepackage{epstopdf}

\usepackage[font=small,skip=0pt]{caption}
\usepackage[numbers,sort&compress,square]{natbib}
\makenomenclature

\usepackage{subcaption}
\usepackage{graphicx}  
\usepackage{soul}

\def\saeed{\textcolor{blue}}
\def\review{\textcolor{black}}

\begin{document}

\title{
Optimal OLTC Voltage Control Scheme to Enable High Solar Penetrations
}

\author{\IEEEauthorblockN{Changfu Li, Vahid R. Disfani, \textit{Member, IEEE}, Zachary K. Pecenak, Saeed Mohajeryami, and Jan Kleissl}

\thanks{Changfu Li, Vahid R. Disfani, Zachary K. Pecenak, Saeed Mohajeryami and Jan Kleissl are with the Center for Energy Research and the Department of Mechanical and Aerospace Engineering at the University of California San Diego, La Jolla, CA 92093 USA. \{emails: chl447@ucsd.edu, disfani@ucsd.edu, zpecenak@ucsd.edu,
samohajeryami@ucsd.edu,
jkleissl@ucsd.edu\}}

\thanks{Vahid R. Disfani is also with the Department of Electrical Engineering at the University of Tennessee at Chattanooga, Chattanooga, TN 37403 USA. \{email: vahid-disfani@utc.edu\}}
}

\maketitle
\IEEEpeerreviewmaketitle
\begin{abstract}


High solar Photovoltaic (PV) penetration on distribution systems can cause over-voltage problems. To this end, an Optimal Tap Control (OTC) method is proposed to regulate On-Load Tap Changers (OLTCs) by minimizing the maximum deviation of the voltage profile from 1~p.u. on the entire feeder. A secondary objective is to reduce the number of tap operations (TOs), which is implemented for the optimization horizon based on voltage forecasts derived from high resolution PV generation forecasts. A linearization technique is applied to make the optimization problem convex and able to be solved at operational timescales. Simulations on a PC show the solution time for one time step is only 1.1~s for a large feeder with 4 OLTCs and 1623 buses. OTC results are compared against existing methods through simulations on two feeders in the Californian network. OTC is firstly compared against an advanced rule-based Voltage Level Control (VLC) method. OTC and VLC achieve the same reduction of voltage violations, but unlike VLC, OTC is capable of coordinating multiple OLTCs. Scalability to multiple OLTCs is therefore demonstrated against a basic conventional rule-based control method called Autonomous Tap Control (ATC). Comparing to ATC, the test feeder under control of OTC can accommodate around 67\% more PV without over-voltage issues. Though a side effect of OTC is an increase in tap operations, the secondary objective functionally balances operations between all OLTCs such that impacts on their lifetime and maintenance are minimized.
\end{abstract}

\nomenclature{$Y$}{Admittance matrix}%
\nomenclature{$Z$}{Impedance matrix}%
\nomenclature{$p$}{Index of OLTC}%
\nomenclature{$t$}{Index of time step}%
\nomenclature{$P$}{Set of all OLTCs}%
\nomenclature{$T$}{Set of all time steps}%
\nomenclature{$a$}{Tap ratio}%
\nomenclature{$\tau$}{Tap position}%
\nomenclature{$J$}{Objective function}%
\nomenclature{$w_1$}{Weighting factor of voltage deviation objective $J_1$}%
\nomenclature{$w_2$}{Weighting factor of tap operation objective $J_2$}%
\nomenclature{ATC}{Autonomous tap control}%
\nomenclature{OLTC}{On-load tap changer}%
\nomenclature{PV}{Photovoltaic}%
\nomenclature{TO}{Tap operation}%
\nomenclature{VLC}{Voltage level control}%
\nomenclature{OTC}{Optimal tap control}%
\nomenclature{VDG}{Variable distributed generation}%
\nomenclature{DSTATCOM}{Distribution static compensator}%
\nomenclature{SVC}{Static Var Compensator}%
\nomenclature{SC}{Shunt capacitor}%
\nomenclature{VAr}{Volt-ampere reactive}
\nomenclature{ShR}{Shunt reactor}
\nomenclature{DSO}{Distribution system operator}

\printnomenclature

\section{Introduction}

The amount of variable distributed generation (VDG) such as solar PV being connected to the grid continues to increase each year as a result of their many technical, economic, and environmental benefits \cite{Walling2008}. However, existing distribution networks may not be capable of handling large amounts of VDG since they are initially designed assuming centralized off-site generation.

Variability and intermittency of VDGs, in particular, present significant challenges to voltage regulation in distribution systems \cite{pecenak2017smart}. Traditionally, to address the voltage issues on the distribution system, OLTCs and voltage regulators are typically employed to maintain the voltage on the secondary side of power transformers within regulatory limits 

Conventional ATC of OLTC maintains a fixed voltage at the transformer's secondary side based on measured local busbar voltage, a line-drop compensator, or remote voltage measurements \cite{caldon2013simplified}. As OLTCs are typically configured assuming a voltage drop along the feeder, a voltage rise caused by reverse power flow during periods with low demand and high solar power feed-in can lead to overvoltages \cite{carvalho2008distributed}.
Moreover, with high PV penetration on a distribution system, high frequency solar ramping caused by fast-moving clouds can result in excessive TOs \cite{Yan2012}.

In order to solve voltage problems resulting from high penetration of VDG on distribution systems, various advanced control methods have been proposed. Several researchers applied rule-based control of OLTCs by replacing local busbar voltage with voltage measurements or estimates from feeder end points and/or critical nodes as the control signal \cite{Stifter2011,Schwalbe2013,Procopiou2016,caldon2013simplified}. However, in these works, the voltage measurements/estimates are only used to control tap position of substation OLTC and coordination between multiple OLTCs is not studied. Therefore, these methods suffer from lack of scalability and are not applicable to feeders with multiple OLTCs.

Other researchers exploited the capability of devices other than OLTCs. A DSTATCOM was used in \cite{yan2014impacts} to damp impacts of residential PV power fluctuations on the OLTC operation. However, the study is done on a small balanced network and no coordination is considered between DSTATCOM and OLTCs. Coordination between energy storage systems and OLTC is studied in \cite{Liu2012} for peak load shaving, power loss reduction, and tap changer stress relief (reduction in TO and reducing operations close to tap limits).
However, the proposed solution requires adoption of costly battery storage systems and does not consider coordination of multiple OLTCs. The authors in \cite{AbdelRahman2006} have studied voltage control by using faster static var compensator (SVC) and slower-responding OLTC to limit SVC reactive power output and reset voltage reference after disturbances for effective voltage support. However, the control is only based on local voltage measurements and as a result the SVC and OLTC are not truly coordinated in an optimal way. Moreover, only one OLTC is used in this work, which makes the scalability of the proposed approach questionable.

A central control methodology can achieve coordinated control of different voltage regulation devices and optimize their operations over a time horizon by taking advantage of load and solar forecast. In \cite{Senjyu2008}, OLTCs, shunt capacitors (SCs), shunt reactors (ShRs), and SVC are optimally dispatched hourly to minimize voltage deviation and energy losses. Similarly, reference \cite{Oshiro2011}
updates optimal tap position of OLTC and reactive power output of PV inverters every 50~s to minimize voltage deviation. Simulation results in both studies show that the proposed methods are able to achieve the desired objectives. However in both works, a genetic algorithm is used to search for the optimal solution, which can be time-consuming considering the extensive search space for the coordinated control of different devices.

Reference \cite{Daratha2014} presents a two-stage approach for solving the optimal voltage regulation problem with coordination between OLTC and SVC. The optimization problem is solved hourly to minimize power losses and TO based on one hour ahead forecasts. The two-stage method is also adopted to solve the coordination problem of more devices including OLTC, SC, and SVC under load and distributed generation uncertainty in \cite{Daratha2015}. Although the proposed method already improves the solution time by a large margin comparing to existing methods, it still takes around 25~s to solve the problem for one time step on the small IEEE 123-bus system in \cite{Daratha2014} and the solution time increases to 58~s in \cite{Daratha2015} which has more devices on the same test feeder. Since PV variability in partly cloudy conditions over a distribution system typically occurs on the order of a few minutes, an hourly time step is insufficient. Rather sub-minute time steps are recommended and therefore this two-stage approach is still questionable for high-resolution application for real distribution feeders, which usually contain thousands of buses. Reference \cite{Agalgaonkar2014} proposes and successfully demonstrates a coordinated reactive power control of PV to minimize TO and avoid operating the OLTC at its control limits. However, only coordination of PV and substation OLTC is considered, while the two other OLTCs operates autonomously.

Despite showing promising results, all of the optimal control methods in \cite{Senjyu2008, Oshiro2011, Daratha2014, Daratha2015, Agalgaonkar2014} are tested on simple distribution networks with only a few or evenly-distributed PV systems. In terms of PV generation profiles, only reference \cite{Oshiro2011} adopts the required sub-minute generation profile during partly cloudy conditions while low-resolution (1~h) generation profiles for a clear-sky day are used in \cite{Senjyu2008, Daratha2014, Daratha2015}. Reference \cite{Agalgaonkar2014} applies PV profiles with 30~s resolution but it is also for clear day without solar ramps. Morever, the same generation profile is used for all PVs in these studies even though it is very critical to use unique and realistic generation profiles for each PV, and the importance of applying realistic individual PV generation profiles has been demonstrated in \cite{Nguyen2016}.

In addition, TO step limits between two consecutive simulation time steps have not been considered which could lead to unrealistic operating decisions of OLTC. For example,  reference \cite{Daratha2014} provides an additional test case to show the proposed method's ability of dealing with fast-moving clouds effects, however, the results shows that one OLTC would need to switch by eight steps in less than 1 minute, which is a challenging and arguably impractical task for conventional OLTC with slow mechanical switching gear and the typical 30 to 60~s time delay \cite{loadtapchangecontrol}.


In summary, application of existing optimal control methods to sub-minute high-resolution applications are questionable. Potential issues include large computation time, lack of consideration of realistic OLTC switch limits, limited testing on large real distribution feeder and realistic representation of distributed PV characteristics like random deployments and fast ramping events.
To tackle these issues, we propose a multi-horizon, central optimization of OLTC tap position to minimize voltage deviation maxima throughout the feeder and minimize the number of TOs.
High temporal and spatial resolution PV forecasts are employed to reflect a realistic picture of a feeder with high penetration of distributed VDGs.

The contributions of this paper to improve the state-of-the-art in optimal voltage control are:
\begin{enumerate}
\item A novel linearization technique to represent the OLTC tap position and feeder voltage as a convex optimization problem which is
solvable in an operational time scale at high-resolution (30~s).
\item Coordination between multiple OLTCs 
\item A flexible optimization platform that can be easily expanded to consider other optimization objectives and coordinated control of other devices including SCs and ShRs.
\item Demonstration of the method in realistic conditions on real feeders.
\item Using realistic OLTC switching limits.
\end{enumerate}

The proposed method is firstly benchmarked against an advanced rule-based control method that is found in the literature and proven to be effective through simulations and field deployments. Since the rule-based control method does not coordinate multiple OLTCs, the proposed method is further compared against a conventional autonomous control method. These studies are carried out through simulations on two disparate California distribution feeders.

The rest of the paper is organized as follows. Section \ref{LMTOVI} discusses the tap operation and their effects on feeder voltages. Section \ref{FWOTCC} introduces the optimization platform. Section \ref{casestudy} provides details of the test feeder models, control concepts, and simulation scenarios. Section \ref{result} presents simulation results followed by conclusions and future work in section \ref{Conc}.

\section{Linearized Model of Tap Operation Voltage Impacts}
\label{LMTOVI}


\subsection{Voltage Effects of OLTC Tap Operation}
\label{VoltEffectTO}
OLTCs are indispensable in regulating voltage. They include a moving connection point (called tap) along a transformer winding which allows discrete numbers of turns to be selected. Along the transformer winding, the residing points for taps are called tap positions and denoted by $\tau$. OLTCs regulate voltage by altering the tap position and thus changing the ratio of secondary voltage with respect to the primary voltage. The ratio is referred as tap ratio $a$. The tap positions are numbered such that $a=1$ when $\tau=0$.


\review{The net node current injections of any distribution feeder ($I$) can be represented by the following equation,}
\begin{align}
I=YV, \label{I-V}
\end{align}
where $Y$ is the admittance matrix of the feeder and $V$ is the complex vector of node voltages. Similarly, the vector of node voltages can be written as a function of injected currents as
\begin{align}
V=ZI,
\label{V-I}
\end{align}
where $Z=Y^{-1}$ is the feeder impedance matrix.
 A linear approximation of the perturbations in node voltage due to changes in impedance and current ($\partial V / \partial (ZI)$) leads to
\begin{align}
\Delta V=\Delta Z\cdot I+Z \cdot \Delta I.
\label{VI_derivaive}
\end{align}

To make the problem mathematically tractable, it is assumed that $\Delta I = 0$, and with this assumption the second term on the RHS of Eq.~\eqref{VI_derivaive} will become zero. $\Delta I = 0$ means that the current injection is fixed at each time step as used in \cite{Pecenak2017}. The reason behind this assumption is that loads and PVs, i.e. current sources, are not controlled in this optimization problem. Even if their current injections changes, they are negligible and due to voltage change caused by tap changes from initial tap positions. In Section \ref{acc}, a sensitivity analysis is performed to verify the accuracy of this assumption, and it is shown that this assumption is valid.

To proceed, a model to determine the impacts of TO on $\Delta Z$ is required. We propose a novel method to modeling effects of tap position on $\Delta Z$. Assuming a OLTC is connected between the node $i$ of the primary side and node $j$ of the secondary side, a change in the tap position of OLTC affects only the elements of the admittance matrix corresponding to these two nodes as
\begin{align}
&Y_{ii}=a^2/z_T+\sum_{k\neq j}{1/z_{ik}}  \label{Yii}\\
&Y_{ji}=Y_{ij}=-a/z_T \label{Yij}  \\
&Y_{jj}=1/z_T+\sum_{k\neq i}{1/z_{jk}}, \label{Yjj}
\end{align}
where $z_T$ is the equivalent impedance of the transformer on the winding connected to node $i$, $z_{ik}$ is the impedance connected between two arbitrary nodes $i$ and $k$ ($i\neq k$), and $z_{ii}$ is the impedance connected from the node $i$ to ground.

Let us define $Y_0$ and $Z_0=Y_0^{-1}$
as the admittance and impedance matrices for the initial tap ratio $a_0$. Similarly, $V_0$ and $I_0=Y_0V_0$ denote the resulting node voltage and injected current for $a_0$.  According to Eq.~\eqref{I-V}, any change in injected current can be represented as
\begin{align}
\Delta I=Y_0\cdot\Delta V + \Delta Y\cdot V_0.
\label{DI0}
\end{align}

Under the assumption of fixed current injection from loads and PVs, change of current injections are equal to zero. Therefore,
Eq.~\eqref{DI0} is equal to zero yielding,
\begin{align}
\Delta V=-Y_0^{-1}\cdot\Delta Y\cdot V_0
\label{DeltaV}
\end{align}

Similarly, since $\Delta I$ equals to zero, Eq.~\eqref{VI_derivaive} can be written as,
\begin{align}
\Delta V=\Delta Z\cdot I_0.
\label{DeltaV_}
\end{align}

Combining Eq.~\eqref{DeltaV} and Eq.~\eqref{DeltaV_}, the matrix $\Delta Z$ becomes
\begin{align}
\Delta Z=-Y_0^{-1}\cdot\Delta Y\cdot Y_0^{-1},
\label{DeltaZ}
\end{align}
which is input to Eq.~\eqref{DeltaV_} to determine $\Delta V$.

\vspace{-2ex}
\subsection{Linearization of $\Delta Y$}

From Eq.~\eqref{DeltaV_}, $\Delta V$ can be determined if $\Delta Z$ is known. Therefore, $\Delta Z$ is used for voltage control in the optimization. And the control of $\Delta Z$ is achieved by control of $\Delta Y$ via changing tap position. We design the voltage control platform as a convex optimization problem to reduce computational expense and achieve global optimality. To achieve convexity, the objective function and optimization constraints have to be linearized.

The admittance matrix depends non-linearly on tap position as some elements {are a function of} $a^2$ (see Eq.~\eqref{Yii}). A tap change from $a_0$ to $a$ yields $\Delta Y_{ii} = (a^2-a_{0}^2)/z_T$. To remove the non-linearity, a Taylor series expansion is performed for $a^2$ around $a_0$. Replacing $a^2$ with $2aa_0 - a_0^2$ yields a linear expression $\Delta Y_{ii} = (2aa_0-2a_{0}^2)/z_T$.

\subsection{Linearization of voltage magnitudes}
Node voltages are the main control parameters in the voltage control algorithm. \review{Magnitude of node voltages expressed as complex numbers can be calculated using their real parts and imaginary parts by the following equation,
\begin{align}
|v|^2=v_d^2+v_q^2,\label{vamag}
\end{align}
where $v=v_d+jv_q$ is the complex voltage of an arbitrary node in the feeder.}

The magnitudes of node voltages in Eq.~\eqref{vamag} are linearized around the operation point (i.e. $v_0=v_{d_0}+jv_{q_0}$) as
\begin{align}
|v_0|\Delta|v|=v_{d_0}\Delta v_d+v_{q_0}\Delta v_q
\label{vamaglin}
\end{align}
and then,
\begin{align}
|v|=|v_0|+\Delta|v|=|v_0|+|v_0|^{-1}(v_{d_0}\Delta v_d+v_{q_0}\Delta v_q).
\label{vamaglin}
\end{align}

This definition for voltage magnitudes of all nodes sets up an affine relation between the voltage magnitude and the OLTC tap position and makes the optimization problem convex.


\section{Feeder-Wide Optimal Tap Changer Control}
\label{FWOTCC}
\subsection{Optimal OLTC Control Structure}


In the future smart grid, smart meters, synchrophasors, and communication platforms enable distribution system operator (DSO) to be aware of the system states (voltage magnitudes and angles) of all buses of the feeder \cite{Farhangi2010}. Moreover, solar and demand forecast can provide insights of future conditions of the grid.

We propose that non-local information could be used by DSO to develop a central or feeder-wide optimal voltage control platform, which can monitor direct effects of OLTC tap position on the voltage profile of the entire feeder.

\begin{figure}
\centering
\includegraphics[width=0.5\textwidth]{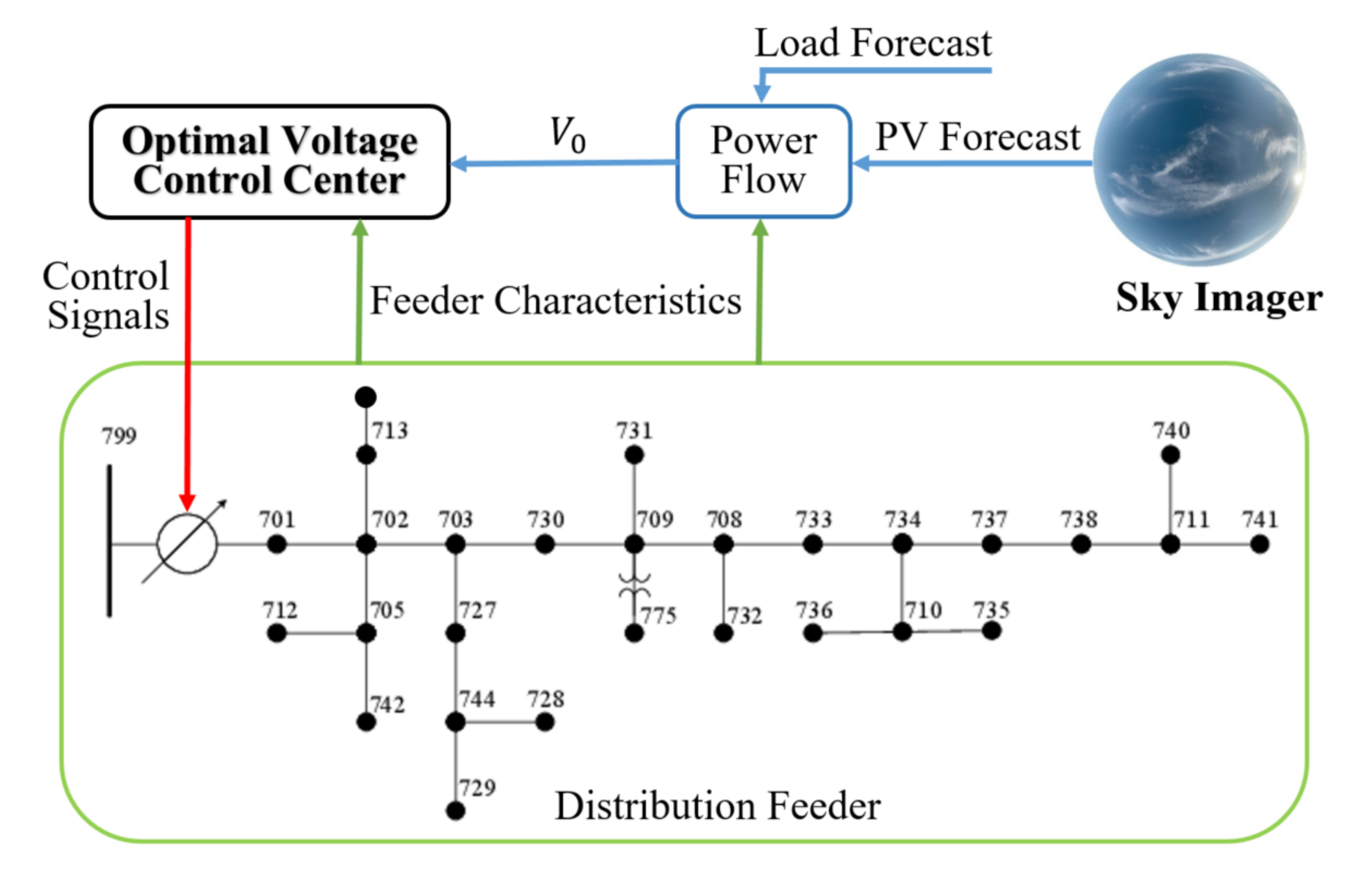}
\caption[Feeder Topologies]{Structure for proposed optimal voltage control.
}
\label{OVCC}
\end{figure}

\review{Fig. \ref{OVCC} shows the central control platform. The control center obtains voltage states ($V_0$) at the linearization point from power flow run by OpenDSS\cite{OpenDSS}. The solution of the optimization problem for the control horizon using PV and load forecasts then yields optimal tap positions of all OLTCs. The optimal tap positions are then delivered back to each OLTC without communication delay. For readability only a single OLTC is shown, but the method can handle any number of OLTCs.}


Due to the discrete nature of tap positions as the main decision variables, the corresponding optimization problem is a mixed-integer programming problem.
To minimize the number of TO in the control objectives, the optimization problem is defined over a 5 min horizon. \review{The 5 min horizon is chosen in accordance with the Sky Imager's forecast horizon, which can provide forecast of PV availability for up to 10~mins The forecast resolution is 30~s. The optimization horizon can assume any value shorter than the forecast horizon.} Details regarding the sky imager forecasting can be found in \cite{Chow2011,yang2014solar}.
\review{Historical measurements provided from the utility are used as perfect load forecast, which is also the practice in \cite{Senjyu2008, Oshiro2011}.}
 The forecasted conditions allow the tap position to be optimized based on present and future grid states.

\vspace{-2ex}
\subsection{Optimization Model}
\subsubsection{Objective Functions}
Since the goal is to improve feeder voltage profile while minimizing tap operations, two objective functions are defined. The first objective function ($J_1$) minimizes the voltage deviations on the feeder during the optimization horizon. The specific parameter selected is the maximum voltage deviation,
\begin{align}
J_1=\max_{t\in T}\left\{\max\left\{||V_t|-\mathbf{1}|\right\}\right\},
\label{J1}
\end{align}
where $T$ is the set of time steps in the optimization horizon, $|V_t|$ denotes the vector of voltage magnitude of all nodes at time step $t$, and $\mathbf{1}$ is a vector with all its components being equal to 1~p.u.. Minimizing the maximum voltage deviation across the feeder from 1 p.u. centers the max and min voltage of the feeder around 1 p.u. \review{to keep the voltage of the entire feeder within the 0.95 to 1.05~p.u. ANSI limits \cite{std2006c84}.}

The second objective function ($J_2$) counts the number of
{TO} as,
\begin{align}
J_2=\sum_{t\in T}\sum_{p\in P}|{\tau_{p,t+1}-\tau_{p,t}}|,
\end{align}
where $P$ is the set of all OLTCs and $\tau_{p,t}$ denotes the tap position of OLTC $p$ at time step $t$. All tap changes over any two consecutive time steps over a defined time horizon $T$ are aggregated in $J_2$.

\subsubsection{Constraints}
 To ensure that the final solution meets the feeder power flow, the linearized power flow (Eq.~\eqref{DeltaV}) is an equality constraint. Further, the equality constraint in Eq.~\eqref{vamaglin} relates the voltage magnitudes to the real and imaginary parts of the node voltages. Tap positions $\tau$ are integer values in the range of $[-\tau_{\max},\tau_{\max}]$. Assuming that the selection of $\tau_{p,\max}$ for OLTC $p$ leads to a tap ratio equal to $a_{p,\max}$,
its tap ratio at time $t$ can be represented as an affine function of $\tau_{p,t}$ as
\begin{align}
a_{p,t}=1+\frac{\tau_{p,t}}{\tau_{p,\max}}(a_{p,\max}-1)
\label{atau}
\end{align}
for any $t\in T$ and $p\in P$. Therefore, Eq.~\eqref{atau} must be included in the optimization model as an equality constraint.

To avoid unrealistic tap operation as reported in \cite{Daratha2014} and to consider TO delays, the optimization problem restricts the number of TO between two consecutive time to be less than $\Delta TO_\text{max}$, which is set to be 1 in the paper.

\subsubsection{Optimization Formulation}
Combining the two objective functions, the optimal voltage control problem is
\begin{align}
\min &&J=w_1J_1+w_2J_2\label{optimization_problem}\\
\text{s.t.}&&\eqref{DeltaV}, \eqref{vamaglin}, \eqref{atau}\nonumber\\
&&\tau_{p,t}\in\mathbb{Z}&&\forall_{p\in P}~\forall_{t\in T}\nonumber\\
&&-\tau_{p,\max}\le\tau_{p,t}\le\tau_{p,\max}&&\forall_{p\in P}~\forall_{t\in T}\nonumber\\
&&|\tau_{p,t}-\tau_{p,t-1}|\le \Delta TO_{p,\max}&&\forall_{p\in P}~\forall_{t\in T}\nonumber
\end{align}
The weighting factors, $w_1$ and $w_2$ are chosen to be 1 and 0.01, respectively. The reason behind the selection of these values are fully elaborated in Section \ref{wf}.

\subsubsection{Proof of Convexity}
\label{convex}
Since absolute values, maximum values, and summation of convex functions are still convex, we claim that the objective functions $J_1$ and $J_2$ and their sum are convex. Moreover, all the constraints are affine functions. These two characteristics preserve the convexity of the optimization problem. Therefore, its convergence to the global optimal solution is guaranteed in polynomial time.

\section{Case Study}
\label{casestudy}

\subsection{Distribution Feeder Models}
\begin{figure}
\centering
   \begin{subfigure}[b]{0.45\textwidth}
  \includegraphics[width=0.99\textwidth]{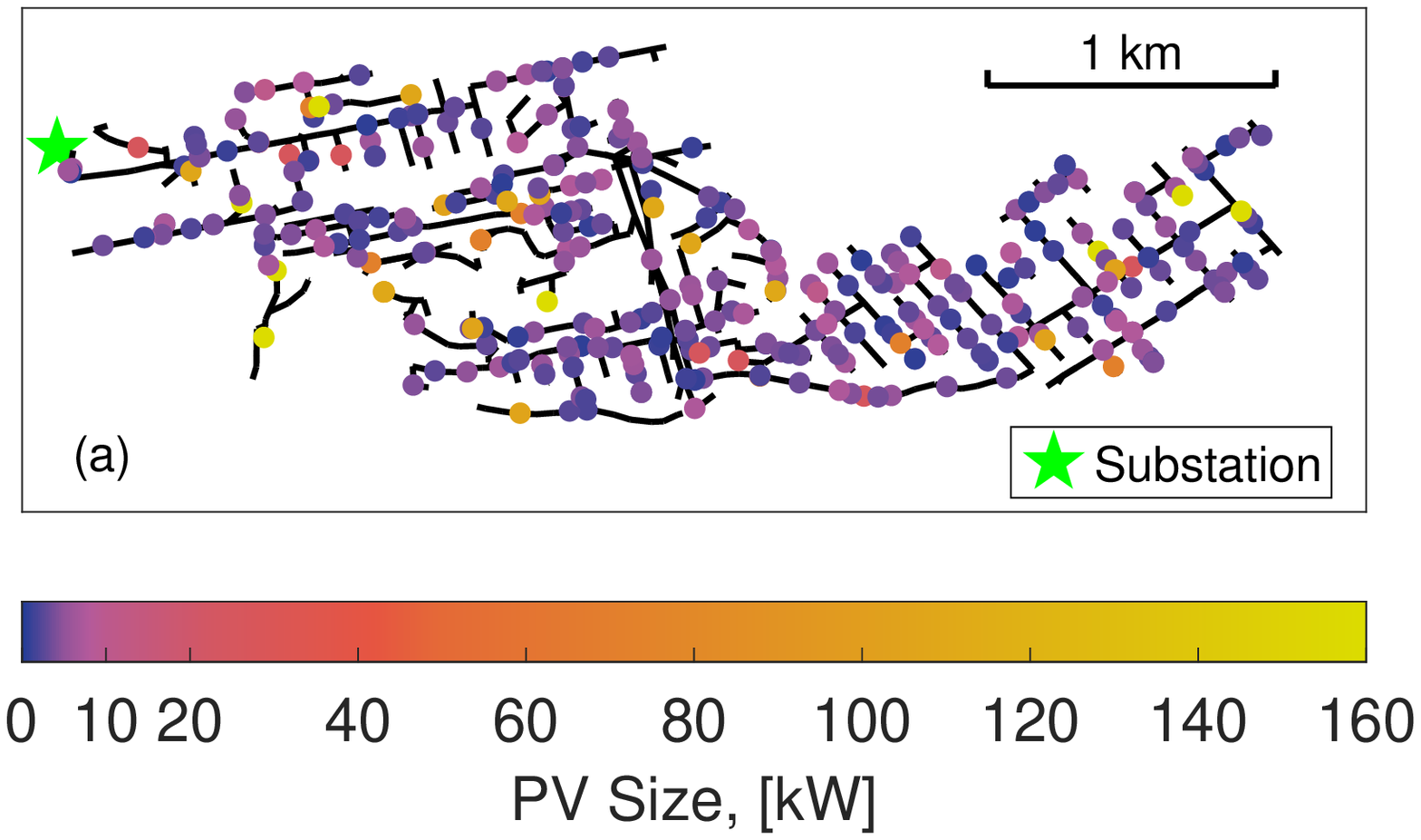}
   \label{topologyA}
\end{subfigure}

\begin{subfigure}[b]{0.50\textwidth}
\includegraphics[width=0.95\textwidth]{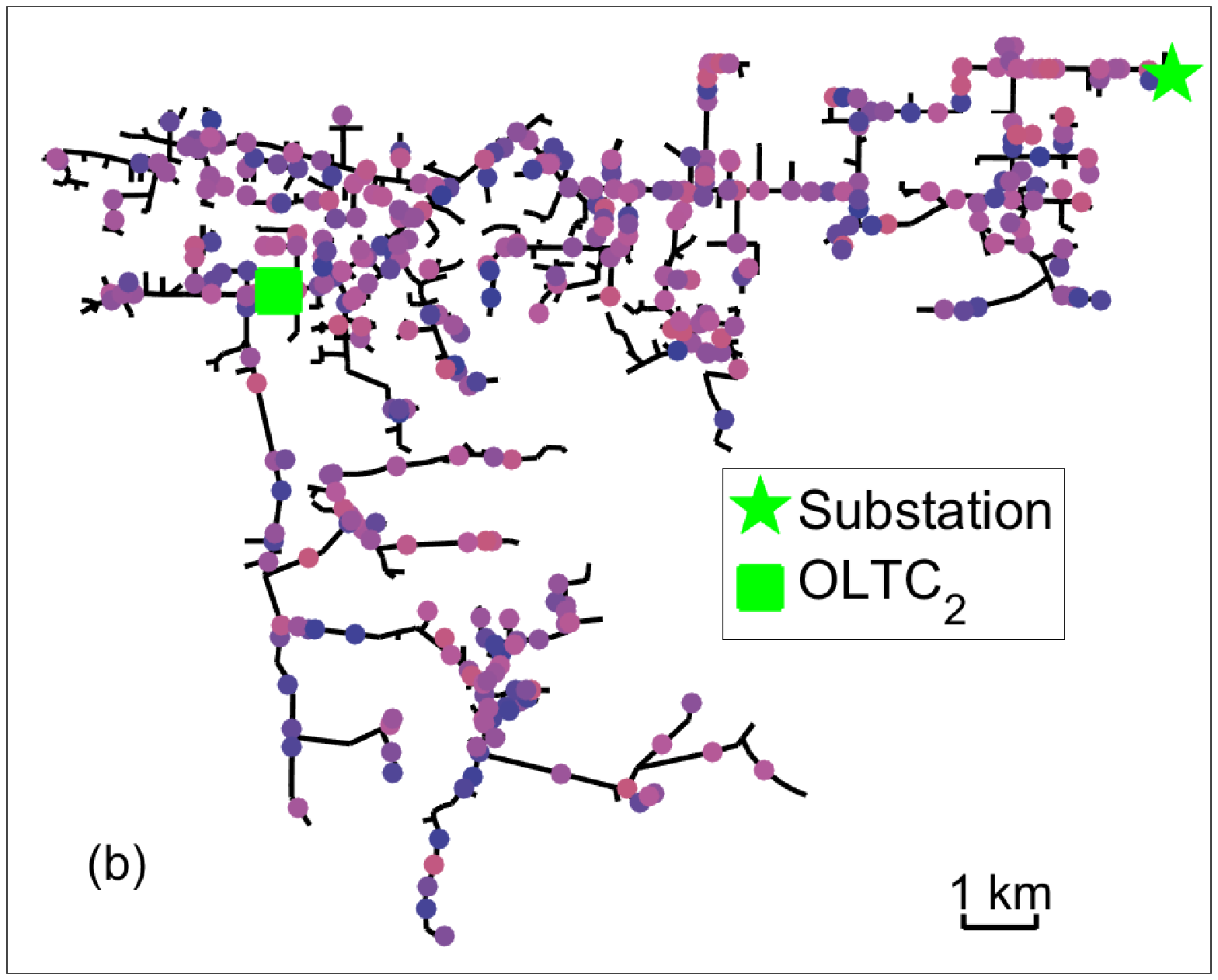}
   \label{topologyB}
\end{subfigure}

\caption[Feeder Topologies]{Feeder topologies of feeder (a) A and (b) B. Black lines represent feeder lines. Each dot is a PV system and its color indicates its AC power rating. The substation is marked with a green star and the green square marks the additional OLTC of feeder B.

\vspace{-2ex}
\vspace{-0ex}}
\label{topology}
\end{figure}

To evaluate the proposed OTC method, quasi-steady state simulations are carried out for two real California distribution feeders.
The feeders, referred to as feeder A and B in this paper, correspond to feeder 2 and feeder 5 in \cite{Nguyen2016}, where more details regarding feeder models, load data, and PV generation data can be found. Feeder topologies of the chosen are displayed in Fig.~\ref{topology}. And table~\ref{feederABtable} summarize feeder characteristics.

Feeder A and B are chosen to represent feeders with a single OLTC and multiple OLTCs, respectively. Feeder A is equipped with a single OLTC at the substation. Feeder B has one OLTC installed at the substation and  a second OLTC located in the middle of the feeder as shown in Fig.~\ref{topology}. Tap position of all OLTCs can vary from -16 to +16 with voltage regulation capability of [0.9 1.1] p.u..

Capacitors are removed from the circuits due to convergence issues at high PV penetration, and thus the feeder under ATC experiences under-voltage problems in the morning and evening independent of PV penetration. Under high PV penetration the maximum voltage on the feeder always occurs in the middle of the day regardless of the tap control method. The capacitors would have little effect (if any) on the voltage in the middle of the day with high PV penetration since the capacitors typically switch off or operate with small VArs at those times. Since the ATC undervoltages would not occur in reality on these two feeders with capacitors, the comparison of the three control methods is only based on over-voltages.

\subsection{Tap Control Schemes}
\label{Def}

In this paper, OTC is firstly benchmarked against an advanced rule-based VLC found in literature. Since the VLC does not coordinate multiple OLTCs, the OTC is then compared with a basic rule-based ATC. Details regarding these three control schemes are provided below.

\subsubsection{Autonomous Tap Control (ATC)} ATC denotes the widely used conventional OLTC operation where the OLTCs only monitor their local busbar voltage and change tap to keep the deviation of the local busbar voltage from the preset reference voltage within certain limits. To avoid excessive TOs, a tap change is only triggered when the measured voltage is out of range for a certain period of time. The minimum time period is called tap time delay. \review{A shorter time delay will provide better voltage regulation but at a cost of more TOs, and vice versa.} In this paper, ATC is only applied to feeder B for comparison with OTC when there are multiple OLTCs. The reference voltages of both OLTCs on feeder B are set to 0.99~p.u. and the voltage regulation bandwidth is 0.0167~p.u.. The tap time delay is set to 60~s \review{based on the utility setting.} All other OLTC parameters are kept as default OpenDSS \cite{OpenDSS} values.

\subsubsection{Voltage Level Control (VLC)} VLC denotes the method adapted from \cite{Stifter2011} which controls OLTCs based on voltage measurements from critical nodes of the feeder. This advanced rule-based control method has proven to be effective in simulation as well as field test \cite{Stifter2011,Schwalbe2013}. When there are over- or under-voltage problems on a feeder, VLC dynamically sets a reference voltage based on voltage measurements using the equation:
\begin{align}
U_{\rm new} = U_{\rm UL} - \frac{\rm VB - Rng}{2}-(u_{\rm max}-U_{\rm old}),
\label{vlc}
\end{align}

\noindent where $U_{\rm new}$ is the new reference voltage. $U_{\rm UL}$ is the voltage upper limit, which is 1.05 p.u., $\rm VB = 0.1$~p.u. is the allowable voltage band which is the difference between voltage upper and lower limits, and $\rm Rng$ is the difference between the highest ($u_{\rm max}$) and lowest ($u_{\rm min}$) measured voltage. The old reference voltage $U_{\rm old}$ is modified by changes in tap position. The relationship of reference voltage and tap position ($\tau$) is: \begin{align}
U_{\rm new} = 1+\frac{(1.1-0.9)\tau}{32} ,
\label{tap_volt}
\end{align}

The initial (at midnight of each day) reference voltage is 1 p.u..

\subsubsection{Optimal Tap Control (OTC)}
\label{OTC}
OTC determines optimal control tap of OLTCs based on the feeder-wide voltage profile. Tap position are the outputs of the optimization problem proposed in Section~\ref{FWOTCC}. OLTCs will follow the optimal tap position schedule to maintain the voltage symmetric about 1~p.u. and a reference voltage therefore does not need to be specified. The number of tap changes is limited to one step per simulation time step (30~s); that is, $\rm \Delta TO_{\rm max}=1$ in Eq.~\eqref{optimization_problem}. The same constraint on tap changes between two consecutive time steps is also considered in ATC and VLC.
\vspace{-2ex}

\subsection{Simulation Setup}

Since VLC is limited to feeders with a single OLTC, feeder A is simulated with VLC and simplified OTC (see Section~\ref{SOWF} for more details) to benchmark OTC against the established rule-based VLC. Meanwhile, feeder B is simulated with ATC and complete OTC to show OTC's ability of coordinating multiple OLTCs to improve the voltage profile.

Feeder A and feeder B are simulated on selected days from the 94 day period spanning December 10, 2014 to March 14, 2015. 28 days are chosen to cover different weather conditions (clear, partly cloudy, overcast) and days when the feeders showed the largest voltages in \cite{distributionhotspot}.

The performance of OTC are evaluated for different amounts of PV generation. Here PV penetration is defined as (\ref{PVpen}).
\begin{align}
PV_{\rm Pen}=\frac{P_{\rm pv\_peak}}{P_{\rm load\_peak}} \times 100\%
\label{PVpen}
\end{align}
where $P_{\rm pv\_peak}$ is the total rated AC power of all PV units\saeed{,} and $P_{\rm load\_peak}$ is the peak feeder load. The desired PV penetration level is achieved by scaling the rated output of existing PV systems. Each feeder is simulated from 0\%  to 200\% PV penetration with 25\% increments.

Both feeders experience very high voltages on March 14, 2015 in \cite{distributionhotspot}, a partly cloudy day with large solar ramps. Therefore March 14, 2015 is used to illustrate voltage and tap change results. Simulations for March 14, 2015 showed that the minimum and maximum voltages at different times of the day occur at just 30 out of the 2,844 nodes of feeder A and 67 out of 4,869 nodes of feeder B. To reduce the computational cost, the objective function $J_1$ considers only the nodes with maximum and minimum voltage deviations and the OLTC nodes.

\vspace{-2ex}

\begin{table}[]
\normalsize
\centering
\caption{
Feeder Properties}
\label{feederABtable}
\begin{tabular}{c|c|cll}

Feeder                & A     & B     &  &  \\ \cline{1-3}
Type                  & Urban & Rural &  &  \\
Voltage Level (kV)     & 12   & 12    &  &  \\
Total Length (km)     & 40   & 115    &  &  \\
Peak Load (MVA)       & 8.4   & 6.3   &  &  \\
\# of Loads & 3761     & 1169     &  &  \\
\# of PV System & 340     & 387     &  &  \\
\# of OLTCs & 1     & 2     &  &  \\
\# of Buses       & 948  & 1623   &  &  \\
\# of Nodes       & 2844  & 4869  &  &  \\ \cline{1-3}
\end{tabular}
\end{table}

\subsection{Selection of Weighting Factors of OTC}
\label{wf}
Since the optimization objective $J$ is a weighted sum of $J_1$ and $J_2$, heavy weighting on $J_1$ will improve the voltage profile at the cost of more TOs and vice versa. Therefore, appropriate weighting factors should be chosen to achieve desired trade-off between voltage profile and number of TOs. Several combinations of weighting factors ($w_1$, $w_2$) are tested with simulations on feeder B on March 14, 2015 with 150\% PV penetration (table \ref{weightfactors}).

 As expected, larger weighting factors ($w_2$) on $J_2$ cause decreases in total TO, while maximum voltages generally increase. When there is no penalty on TO (i.e. $w_2=0$), the maximum voltage on the feeder remains low at 1.049~p.u. as the OLTC moves the taps as often as needed to reduce voltage deviation.

Relative to the case with $w_2=0$, $w_2=0.005$ provide a large reduction in TO without increase in maximum voltage. \review{If $w_2>0.005$, the maximum voltage increases without much TO reduction for the feeder studied. Therefore, $w_2$=0.005 is used hereinafter.}

\review{The desirable combination of w1 and w2 for different feeders may vary due to the preferences of the DSO between better voltage regulation or less TO, different locations of OLTCs, feeder topologies, distribution of PVs, etc. Local adjustments of the weighting factors are therefore recommended.}

\begin{table}[]
\normalsize
\centering
\caption{Case study of different objective function weights $w_2$ on feeder B with 150\% PV penetration on March 14, 2015. $w_1$ is fixed at 1.}
\label{weightfactors}
\begin{tabular}{c|ccc}
\hline
$w_1$                & $w_2$      & Max Volt (p.u.) & Total TO \\ \hline
\multirow{6}{*}{1} & 0 & 1.049          & 1569      \\ \cline{2-4}
                   & 0.005   & 1.049          & 101      \\ \cline{2-4}
                   & 0.01    & 1.055          & 80      \\ \cline{2-4}
                   & 0.02    & 1.059          & 68       \\ \cline{2-4}
                   & 0.04    & 1.156          & 38       \\ \hline
\end{tabular}
\end{table}

\subsection{Sensitivity Analysis}
\label{acc}
Given that predicted node voltages
determine the OLTC tap operations, a sensitivity analysis examines the errors resulting from fixed current injections, the linearization of admittance matrix and voltage magnitude in Eq~\eqref{DeltaV}. Errors are defined as the differences in voltage magnitude calculated from Eq~\eqref{vamaglin} versus the (non-linear) power flow results in OpenDSS:
\begin{align}E(t)_j = V_{\rm calculated}(t)_j - V_{\rm OpenDSS}(t)_j,
\end{align}
where $j$ stands for a node and $t$ stands for a time step. The calculated voltage incorporates all error sources resulted from fixed current injection assumption and linearization of admittance matrix and voltage magnitude.

\begin{figure}
\centering
\includegraphics[width=0.485\textwidth]{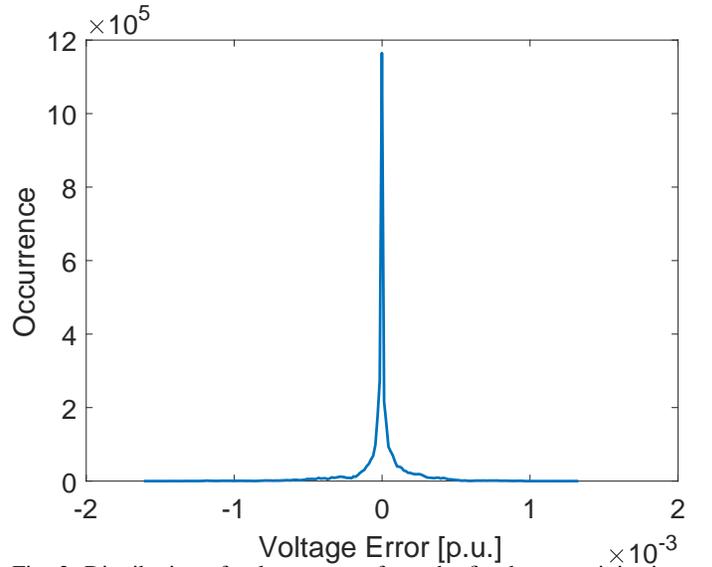}
\caption[]{Distribution of voltage errors from the fixed current injection assumption, linearization of admittance matrix, node voltages  . The distribution is created from simulations for feeder B with 150\% PV penetration.}
\label{volterr}
\end{figure}

Fig.~\ref{volterr} presents error distribution based on simulations on feeder B with 150\% PV penetration. \review{For time steps in the time horizon without tap changes, errors are  nearly identical to zero. Therefore, the figure only shows errors from time horizons with tap changes.} The results show that the calculated voltages almost match the simulated ones. Thus, it can be concluded that the proposed voltage model represents voltage magnitudes very accurately; the maximum error magnitude is 0.0016 p.u. and the mean absolute error magnitude is only \num{9.10e-5} p.u..

\section{Distribution Feeder Simulations Results}
\label{result}

\subsection{Single OLTC without forecasting (Benchmark)}
\label{SOWF}


The OTC approach is benchmarked against the established VLC. Two simplifications are made to OTC for the purpose of making a fair comparison. Firstly, since VLC only changes the tap settings after voltage violations, the simplified OTC platform also changes tap position only if the feeder experiences a voltage violation. In other words the optimal tap position output from OTC is used only if there is voltage violations. Secondly, since VLC can't integrate PV and load forecast to optimize future operations, the OTC control actions are also only based on current conditions. In this way, the basic OTC only calculate tap positions for next time step if needed and therefore TO reduction is not considered. Since both simplified OTC and VLC operate the OLTC to achieve symmetry of maximum and minimum voltage around 1 p.u., OLTC operations under control of both methods are expected to be the same.

\begin{figure}
\centering
\includegraphics[width=0.485\textwidth]{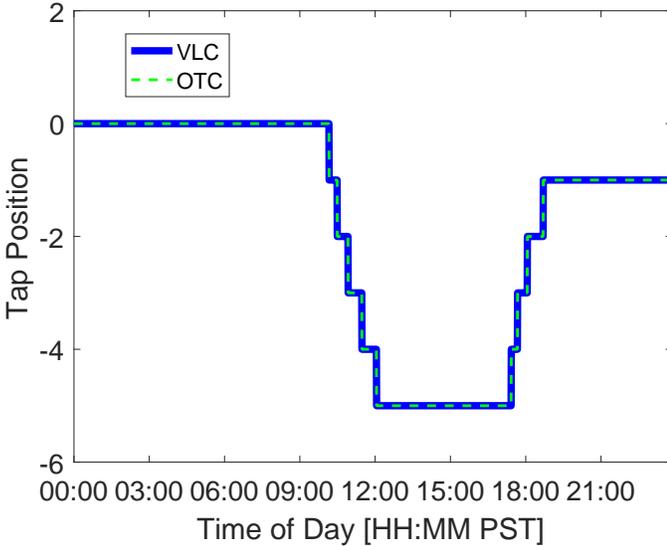}
\caption[]{Time series of the tap position on feeder A with 200\% PV penetration on March 14, 2015.}
\label{to_ltc_otc}
\end{figure}

Fig.~\ref{to_ltc_otc} displays the time series of the OLTC tap position on feeder A under VLC and simplified OTC with 200\% PV penetration on March 14, 2015. The peak output of PV on feeder A is 15.4~MW. The OLTC behaves the same during the whole day under VLC and simplified OTC, resulting in 9 TOs. Therefore, the feeder voltage profiles must also be identical. Simulations for other PV penetration levels show similar results. The OTC and VLC voltages of each node are identical
at all PV penetration levels (0\%--200\%). Therefore, the simplified OTC platform avoids voltage violations as effectively as VLC.

\subsection{Coordination of multiple OLTCs considering forecasting}

To demonstrate OTC's ability of coordinating multiple OLTCs, simulations are carried out for another real California distribution feeder (feeder B) with two OLTCs. Since VLC is unable to coordinate multiple OLTCs, we compare against ATC.

In this comparison, the complete OTC platform described in section~\ref{FWOTCC} is used. OTC optimizes tap position for the next 5~min based on solar and load forecast by balancing the sometimes conflicting objectives of minimum voltage extrema and minimum number of TOs. Further, OTC minimizes voltage deviation extrema even if there is no over- or under-voltage problem.

\subsection*{Voltage Profile}

\begin{figure}
\centering
\includegraphics[width=0.485\textwidth]{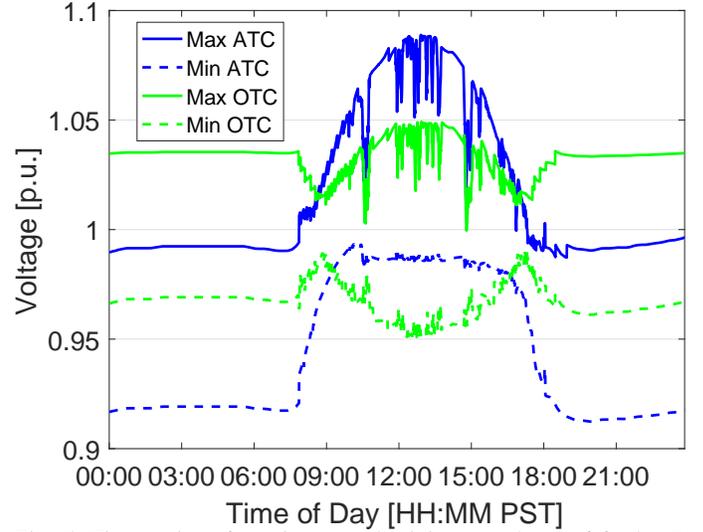}
\caption[]{Time series of maximum and minimum voltage of feeder B with 150\% PV penetration on March 14, 2015.}
\label{volt_ltc_otc}
\end{figure}

Fig. \ref{volt_ltc_otc} presents the time series of maximum and minimum voltage for feeder B with high PV penetration on March 14. The peak output of the PV fleet on the feeder B is 7.6~MW at a time when the load is only 1.8~MW.

The highest and lowest voltages on the feeder are 1.089~p.u. and 0.912~p.u. when the OLTCs are controlled by ATC. With OTC being applied, the highest voltage on feeder A is reduced to 1.049~p.u. and lowest voltage is increased to 0.950~p.u. Although it is expected that maximum and minimum voltage magnitudes deviate symmetrically from 1~p.u., they are slightly asymmetric here (+0.049~p.u. vs. -0.050~p.u.) due to the effect of TO minimization in the objective function and the fact that tap positions are discrete variables.

With ATC, the highest voltage typically occurs near noon when solar irradiance is greatest and the lowest voltage occurs in the evening under high load and low solar irradiance.  With OTC, both the highest and lowest voltages occur in the middle of day. The shift of voltage minima from the evening to the middle of the day is due to the OTC tap position reducing voltage in response to high voltages in the middle of the day. At night the OTC counteracts the voltage drop (caused by increasing load) without support from solar generation, thus the voltage profile is raised. In other words, the minimum and maximum voltages occur at the same time as the largest voltage magnitude change across the feeder. At very high solar penetration the voltage increase across the feeder at midday is larger in magnitude than the voltage drop in the evening.

\begin{figure}
\centering
\includegraphics[width=0.485\textwidth]{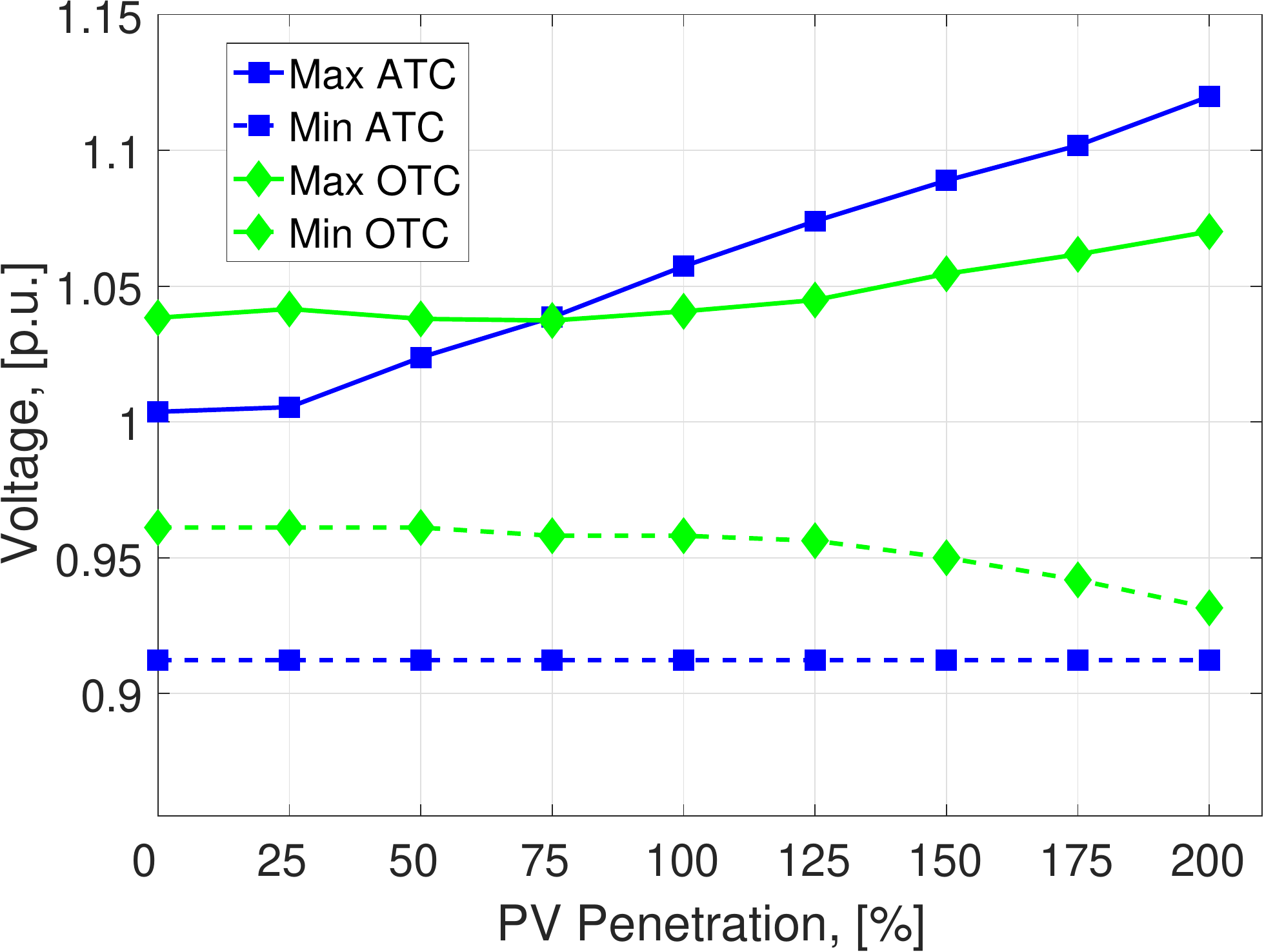}
\caption[]{Maximum and minimum voltage as a function of PV penetration level on March 14 2015 for feeder B.}
\label{volt_vs_pen}
\end{figure}

Fig. \ref{volt_vs_pen} presents the maximum and minimum voltage on the feeder on March 14 as a function of PV penetration level for feeder B. For ATC, maximum voltage increases with PV penetration, while minimum voltage remains constant. The maximum voltages are greater for OTC than ATC at low to moderate PV penetrations (0\%$-$75\%). This is due to the fact that at low PV penetrations OTC corrects for low voltages by raising the tap position. However, at high PV penetration ($>$75\%), OTC maximum voltage becomes lower than for ATC, since OTC reduces over-voltages due to PV generation.

In general, OTC minimizes voltage deviations from 1~p.u., resulting in a voltage profile symmetric about 1~p.u. per the objective function $J_1$,  which is the optimal way to maintain feeder voltage within ANSI limits. Feeder B experiences over-voltage violations starting with around 90\% PV penetration under ATC, while over-voltages do not occur until around 150\% PV penetration with OTC. The OTC scheme therefore allows a 67\% increase in installed PV comparing to the ATC case. The proposed OTC method successfully mitigates over-voltage problems resulting from large amounts of distributed PV generation.

\subsection*{Tap Operations}
\label{TOres}
\begin{figure}
\centering
\includegraphics[width=0.485\textwidth]{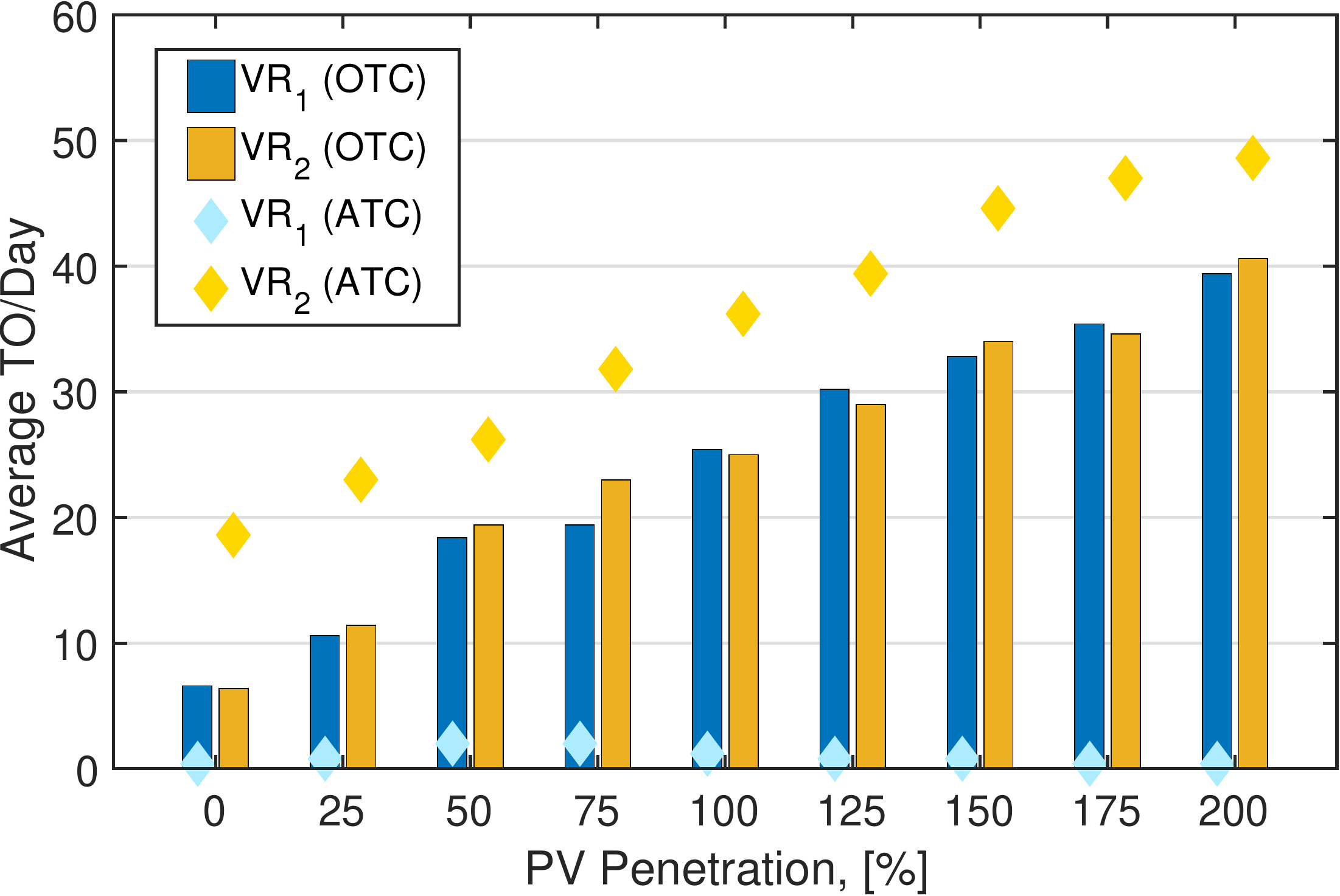}
\caption[voltalong]{Average TO/day of each OLTC on feeder B over the 28~day simulation period as a function of PV penetration.\vspace{-2ex}
}
\label{tovspen}
\end{figure}

The average number of TO per day for each OLTC during the 28~day simulation period for ATC and OTC is shown as a function of PV penetration in Fig.~\ref{tovspen}.  The OLTC at the substation (OLTC$_1$) has low TO for all PV penetrations under ATC while OTC operates OLTC$_1$ more to correct the voltage at the expense of increased TOs. In contrast, OLTC$_2$ generally incurs  about the same TOs with OTC.
TOs could be reduced by applying a larger weight $w_2$ in Eq.~\eqref{optimization_problem} as shown in table \ref{weightfactors}.

OTC optimizes the voltage on the entire feeder and increases TOs of OLTCs located near the substation, while the number of TOs of the downstream OLTC remains around the same. $J_2$ is designed to reduce aggregated TOs of all OLTCs. When OTC detects large voltage deviation on the feeder, it operates OLTCs that are best able to reduce the voltage deviation with a minimum number of TOs. In general, the proposed OTC method can provide a better coordination between OLTCs on the feeder to improve the voltage profile.

Even with PV penetration as high as 200\%, the largest number of TO is observed to be around 50 TO/day for all OLTCs on the feeder. Assuming a lifespan of tap changers of 1 million TO \cite{dohnal2006load}, the OLTC will last for about 55 years. \review{According to \cite{mrload}, the OLTC can perform 600,000 switching operations without maintenance, therefore even the OLTC with maximum TOs under 200\% PV penetration can operate for around 33 years without maintenance.}

\subsection*{Computation Time}

\begin{figure}
\centering
\includegraphics[width=0.485\textwidth]{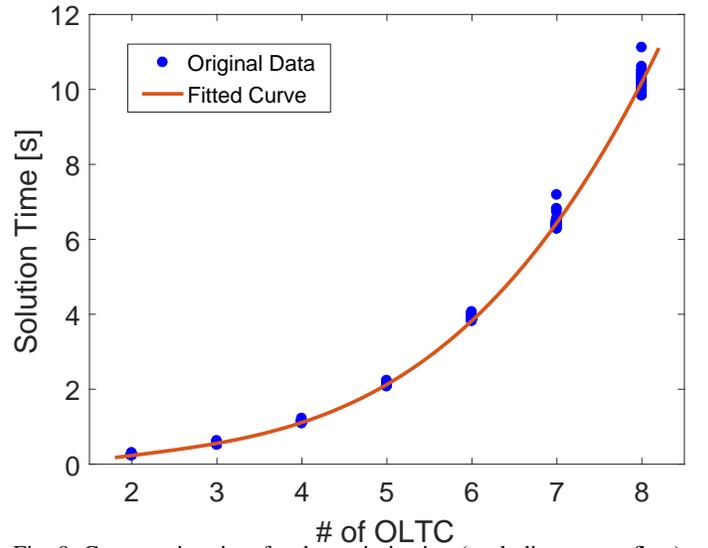}
\caption[]{Computation time for the optimization (excluding power flow) for one time step as a function of total number of OLTCs.\vspace{-2ex}
}
\label{ct}
\end{figure}

Solving the optimization problem at an operational timescale would enable the control to be used in real time applications. To demonstrate the scalability and computational efficiency of proposed OTC, we add up to six OLTCs on feeder B and perform a series of simulations on a PC with an Intel(R) Core(TM) i7-4700MQ 2.8-GHz processor and 16GB RAM.

Fig.~\ref{ct} displays computation time for one time step and two to eight OLTCs together with a curve fit. The expression of the fitted curve is: $y=0.0379\cdot x^3-0.2254\cdot x^2+0.7310\cdot x-0.6309$ (s), where x is the number of OLTC and y is the solution time for one time step. Since the convexity of the optimization problem has been proven in Section \ref{convex}, the solution time of the optimization problem should be polynomial, which is consistent with the fitted expression we get.

The proposed OTC solves the optimization problem efficiently. For a large distribution feeder with 4 OLTCs and 1623 buses the solution time for one simulation time step is 1.1~s . Even with 8 OLTCs, the solution time for one step is still just 10.2~s. In comparison, the method in \cite{Agalgaonkar2014} takes ~25~s to solve the formulated problem for one step on a PC with Intel Xeon E5420 2.5-GHz CPU, 4-GB memory for a feeder with only 4 OLTCs and 123 buses.

\section{conclusions and Future Work}
\label{Conc}

A novel control platform for optimal tap control of OLTC for voltage regulation was proposed. OTC is capable of coordination between multiple OLTCs. OTC is compared against two established rule-based tap control methods through simulations on two real California distribution feeders using high temporal and spacial resolution PV forecast. Results show the OTC can solve the optimization problem in an operational time scale.

Benchmarking showed a simplified OTC to be as effective as an advanced rule-based method in terms of avoiding voltage violations on a feeder with a single OLTC. However, the OTC is more advanced compared to VLC as it considers coordination between two or more OLTCs and integrates forecasts to optimize tap operations.

In the exhibition of the method on a feeder with multiple OLTCs, the complete OTC algorithm outperforms ATC in voltage regulation. Improved voltage regulation enables higher PV hosting capacity on distribution feeders. Comparing to ATC, the feeder tested can accommodate 67\% more PV without over-voltage problem under the control of OTC. This is achieved by operating OLTCs in an effective way that reduces voltage deviation with minimum TOs. In this case, we estimate the life time of the OLTC with most TOs will be around 55 years even with 200\% PV penetration under highly variable solar ramping scenarios. And the OLTC is expected to be able to operate for 33 years without maintenance. 

Future work will include other voltage regulation devices like shunt capacitors, shunt reactors as well as voltage support from PV inverters via reactive power output control. Since OTC relies on accurate solar and power demand forecast to optimize tap positions of OLTCs, its effectiveness of voltage regulation may be compromised by forecasts with low accuracy. The robustness of OTC to deal with  forecast error will be another focus of future work.

\bibliographystyle{IEEEtran}
\bibliography{OptTap.bib}

\end{document}